\documentclass{epl2}

\title{Trapping in complex networks}
\shorttitle{Trapping in complex networks} 

\def\av#1{\left\langle#1\right\rangle}

\usepackage{amsmath,amstext,amssymb,amsfonts}
\usepackage{epsfig}

\author{A. Kittas\inst{1}, S. Carmi\inst{2,3}, S. Havlin\inst{2} \and P. Argyrakis\inst{1}}
\shortauthor{A. Kittas \etal}

\institute{
  \inst{1} Department of Physics, University of Thessaloniki - 54124 Thessaloniki, Greece \\
  \inst{2} Minerva Center \& Department of Physics, Bar-Ilan University - 52900 Ramat Gan, Israel \\
  \inst{3} Center for Polymer Studies, Boston University - Boston, MA 02215 USA}

\abstract{We investigate the trapping problem in Erdos-Renyi (ER)
and Scale-Free (SF) networks. We calculate the evolution of the
particle density $\rho(t)$ of random walkers in the presence of one
or multiple traps with concentration $c$. We show using theory and
simulations that in ER networks, while for short times $\rho(t)
\propto \exp(-Act)$, for longer times $\rho(t)$ exhibits a more
complex behavior, with explicit dependence on both the number of
traps and the size of the network. In SF networks we reveal the
significant impact of the trap's location: $\rho(t)$ is drastically
different when a trap is placed on a random node compared to the
case of the trap being on the node with the maximum connectivity.
For the latter case we find
$\rho(t)\propto\exp\left[-At/N^\frac{\gamma-2}{\gamma-1}\av{k}\right]$
for all $\gamma>2$, where $\gamma$ is the exponent of the degree
distribution $P(k)\propto k^{-\gamma}$.}

\begin{document}

\maketitle

\section{Introduction}

The properties of random walk greatly vary depending on the
dimension and the structure of the medium in which it is confined
\cite{DbA_book,Weiss_book,Redner_book,disorderd_media}, where a
particularly interesting medium for the study of the random walk is
complex networks \cite{gallos1,jasch1,gallos3,Noh,Korean}. Networks
describe systems from various fields, such as communication (e.g.
the Internet), the social sciences, transportation, biology, and
others. Many of these networks are scale-free (SF)
\cite{Barabasi_scaling,BA_review,DM_book,PV_book}. This class of
networks is defined by a broad degree distribution, such as a power
law $P(k)\propto k^{-\gamma}$ ($k \geq m$), where $\gamma$ is a
parameter which controls the broadness of the distribution.

Trapping is a random walk problem in which traps are placed in
random locations, absorbing all walkers that visit them. This
problem was shown to yield different results over different
geometries, dimensions and time regimes
\cite{Weiss_book,hollander,Redner_book,sphere,trap_dist,trap_math}.
The main property of interest during such a process is the survival
probability $\rho(t)$, which denotes the probability that a particle
survives after $t$ steps. The problem was studied in regular
lattices and in fractal spaces
\cite{rosen,Weiss_book,hollander,sphere,trap_math,trap_dist,trap_dist_2} and recently, in
small-world networks \cite{jasch1}.

In this Letter we study the problem of trapping in networks. This is
a model for the propagation of information in certain communication
networks. This follows since in some cases data packets traverse the
network in a random fashion (for example, in wireless sensor
networks \cite{RW_sensor1}, ad-hoc networks \cite{RW_adhoc1} and
peer-to-peer networks \cite{RW_P2P}). A malfunctioning node in which
information is lost (e.g., a router which cannot transmit data due
to some failure) acts just like a trap in the model. This model
can also be applied to loss of information in messages
over communication systems, e.g. in the case of e-mail messages, where a malfunctioning e-mail server
acts as a node absorbing, but not transmitting, all e-mail messages it receives. Furthermore, our model
may be relevant in social systems, where some information may initially spread randomly,
but in later stages it might be held by certain individuals.

We study the survival probability $\rho(t)$ of random walkers on
random regular networks (networks in which all nodes have equal
degree), Erdos-Renyi (ER) networks (a simple model for random
networks in which all links exist with the same probability
\cite{Bollobas,ER1,ER3}), and SF networks. We derive analytical
expressions for $\rho(t)$ for a wide range of trap degrees and
concentrations and highlight the role of the network structure,
obtaining new scaling relations for the survival probability and
average trapping time which are absent in lattices. Our analytical
predictions are confirmed with Monte-Carlo simulations.

\section{Methods}

To perform Monte-Carlo simulations, we generate ER networks by
considering all pairs of nodes and linking a pair with probability
$p$. The construction of an SF network follows the Molloy - Reed
scheme \cite{Molloy_Reed}. Each node $i$ is assigned a number of
links taken from the distribution $P(k)\propto k^{-\gamma}$ and then
open links are connected randomly. The value of $k$ is taken to be
between $m$ (typically 1-3) to $k_{max}=N-1$ (no upper cutoff value
is imposed). We find the largest cluster by using depth-first search
\cite{Coreman} and then discard all nodes that are not in the
largest cluster. Starting from a fixed density of particles
initially placed in random nodes, particles hop with equal
probability to one of their nearest neighbors. Certain nodes are
randomly chosen to serve as traps. These are perfect traps; if a
particle falls on it then it is trapped and removed from the
network. In the case of multiple traps, $n=cN$ traps are placed in
the network, where $c$ is the trap concentration.

\section{Results}

Assume the network has $N$ nodes, average degree $\av{k}$ and $n$
traps. How does $\rho$ change as $t$ increases to $t+1$ (i.e., after
each particle has moved once)? Denote the traps by
$(i_1,i_2,...i_n)$ and define $k_n\equiv k_{i_1}+k_{i_2} + ... +
k_{i_n}$ as the total number of links emanating from all traps. If
at time $t$ a given particle is not on a trap, but will hop on any
of these $k_n$ links on its next step, it will be trapped at time
$t+1$. We approximate the probability for the particle to hop on any
of these $k_n$ links to be proportional to their relative number in
the network, that is, $A\frac{k_n}{N\av{k}}$, where $N\av{k}$ is the
total number of links in the network, and $A=O(1)$ is the proportion
constant which we will study later. In continuous time, this results
in the equation:
\begin{equation}
\frac{d\rho}{dt} = -\rho A\frac{k_n}{N\av{k}}
\end{equation}
whose solution is:
\begin{equation}
\label{rho} \rho(t) = \rho_0\exp[-Atk_n/(N\av{k})]
\end{equation}

Surprisingly, although Eq. (\ref{rho}) is based on a rather simple
approximation, we show below that it predicts very accurately the
survival probability for various network models, time scales, and
trap concentrations. In fact, Eq. (\ref{rho}) can be seen as a
special case of the theory developed earlier in
\cite{sphere,Havlin84,trap_math,grass}, where it was shown, that for a
$d$-dimensional lattice, the survival probability decays as a
stretched exponential $\rho \sim e^{-\alpha t^{\beta}}$ with
$\beta=\frac{d}{d+2}$. Since networks have infinite dimension,
$d/(d+2)\rightarrow 1$ to recover the exponential decay we predict.
Note that the average time before trapping is ${\cal O}(N)$, as
expected from first passage time considerations \cite{redner_dba}
(see also below).

A necessary condition for the above approximation to hold is that
the number of links between the traps is negligible. For ER networks
where links exist independently of one another the probability that
all links emanating from the traps connect to non-traps nodes is
$\left[(N-n)/(N-1)\right]^{k_n}$. Since $\av{k_n}=n\av{k}$, and
$\left[1-n/(N-1)\right]^{n\av{k}} \approx 1-n^2\av{k}/N$ (for $n\ll
N$), we expect that as long as $n \ll \sqrt{N/\av{k}}$, this
condition is satisfied.

In the following, we will apply Eq. (\ref{rho}) to specific network
topologies. In random regular networks, where each node has exactly
$k$ neighbors, we use Eq. (\ref{rho}) by substituting $\av{k}=k$ and
$k_n=nk$:
\begin{equation}
\label{regular} \rho = \exp\left[-Ant/N\right] =
\exp\left[-Act\right]~~(\mbox{Regular, \emph{n} traps}),
\end{equation}
(without loss of generality, we set $\rho(0)=1$). For other networks
one has to take into account the distribution of degrees. Thus, in
order to average $\rho(t)$ over all networks in the ensemble, we
need to condition $\rho$ on $k_n$:
\begin{equation}
\rho(t)_{\rm{net}}=\sum_{k}P\{k_n=k\}\exp[-Atk/(N\av{k})].
\end{equation}
Consider ER networks with one trap ($n=1$):
$P\{k_n=k\}=e^{-\av{k}}\av{k}^k/k!$ is the degree distribution (a
Poisson) \cite{Bollobas,ER1,ER3}. Thus:
\begin{align}
\label{rho_ER_1}
\rho(t)=\sum_{k=1}^{\infty}e^{-\av{k}}\frac{\av{k}^k}{k!}\exp\left[-\frac{Atk}{N\av{k}}\right] \notag \\
=e^{-\av{k}}\sum_{k=1}^{\infty}\frac{\left[\av{k}\exp\left[-\frac{At}{N\av{k}}\right]\right]^k}{k!} \notag \\
\approx\exp\left[-\av{k}\left(1-\exp\left[-\frac{At}{N\av{k}}\right]\right)\right]\\
(\mbox{ER, one trap, random \emph{k}}) \notag
\end{align}
where we start the summation from $k=1$, since we do not place a
trap on an isolated node ($k=0$). However, when evaluating the sum,
we assume the probability for $k=0$ is negligible, which is
justified whenever $\av{k}$ is large enough, which we assume
henceforth. Also, in the simulations, we consider only the largest
connected cluster, which by definition contains no isolated nodes.

\begin{figure*}
\begin{center}
\includegraphics[width=16.5cm]{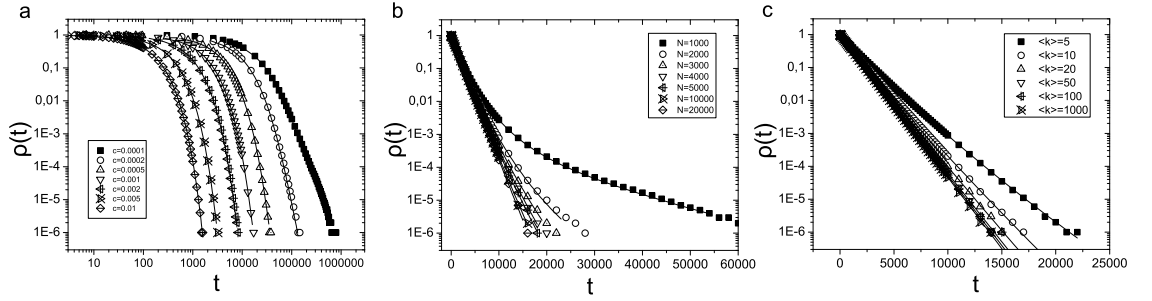}
\caption{Trapping in ER networks. (a) Particle density $\rho(t)$ vs.
$t$ (measured in Monte-Carlo steps).  The network parameters are:
$N=10000$ and $\av{k}=10$. Traps are placed with a concentration $c$
on random nodes of the network. All results are averaged on at least
5000 runs, each with a different configuration of the network. Solid
lines represent fitting with Eq.~(\ref{rho_ER_n}) (with the number
of traps $n=cN$). (b) $\rho(t)$ for fixed trap concentration
$c=0.001$, average degree $\av{k}=10$, and different system sizes.
(c) $\rho(t)$ for fixed trap concentration $c=0.001$, system size
$N=10000$, and different average degrees.} \label{fig1}
\end{center}
\end{figure*}

The same approach can be applied to the case of multiple traps, by
realizing that (neglecting links between the traps) the sum of links
emanating from the traps is a sum of Poisson variables with mean
$\av{k}$, which is itself a Poisson with mean $n\av{k}$:
\begin{align}
\label{rho_ER_n}
\rho(t)=\sum_{k=n}^{\infty}e^{-n\av{k}}\frac{(n\av{k})^k}{k!}\exp\left[-\frac{Atk}{N\av{k}}\right] \notag \\
\approx\exp\left[-n\av{k}\left(1-\exp\left[-\frac{At}{N\av{k}}\right]\right)\right] \\
(\mbox{ER, \emph{n} traps, random \emph{k}}) \notag
\end{align}
The agreement of Eq. (\ref{rho_ER_n}) with simulation results is
evident from Figure \ref{fig1}. Note that the survival probability
in Eq. (\ref{rho_ER_n}) does not solely depend on the trap
concentration $c \equiv n/N$, but on both $n$ and $N$, except for
the short time limit ($t\ll N\av{k}$), when
$1-\exp\left[-\frac{At}{N\av{k}}\right]\approx\frac{At}{N\av{k}}$
and $\rho\approx\exp\left[-Ant/N\right]=\exp\left[-Act\right]$. For
long times ($t\gg N\av{k}$), due to the exponential dependence on
$t$, the main contribution to the survival probability comes from
configurations in which $k_n$ is small, the probability of which
depends on $n$ alone. On the other hand, the probability that the
particle falls into the trap still depends on the total number of
links $N\av{k}$. Thus, the survival probability depends on both $n$
and $N$ independently. It can also be seen that particles survive
longer as the network becomes smaller (Figure \ref{fig1}(b)) and
sparser (Figure \ref{fig1}(c)).

\begin{figure}
\includegraphics[width=7.5cm]{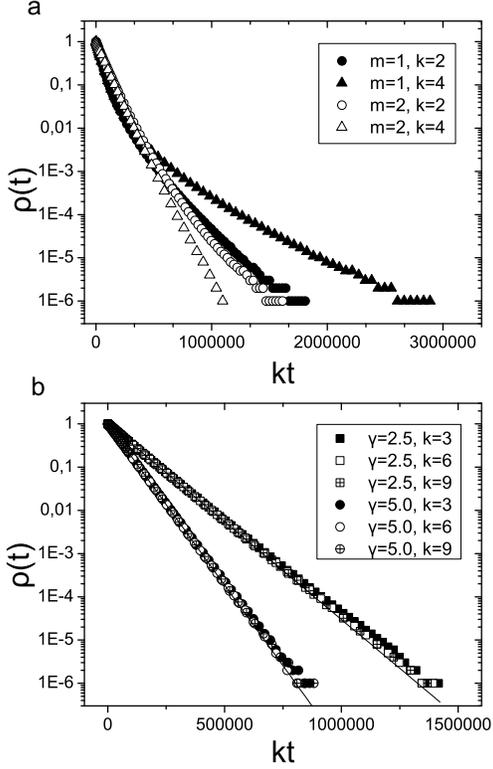}
\caption {Trapping in scale-free networks with a single trap on a
node with fixed degree $k$. (a) Particle density $\rho(t)$ vs. $kt$,
for SF networks with $N=10000$, $\gamma=2.5$, $m=1,2$, and different
trap degrees. (b) Same as (a), but for $m=3$ ($\gamma=2.5,5$). In
this case, all curves collapse, in agreement with Eq.
(\ref{sf_fixed_k}).} \label{fig2}
\end{figure}

Even though scale-free networks are highly heterogeneous and thus
the approximate approach is expected to yield less accurate results,
nevertheless it is still quite useful. The degree distribution is $P(k)= Ck^{-\gamma},
k\geq m$, where $C$ is a normalization factor. Thus, for a single
trap:
\begin{equation}
\rho=\sum_{k}Ck^{-\gamma}\exp\left[-Atk/(N\av{k})\right].
\end{equation}
Since this does not lead to a closed form formula, we focus on the
case where the degree of the trap $k$ is fixed. We expect:
\begin{equation}
\label{sf_fixed_k} \rho=\exp\left[-Akt/(N\av{k})\right]~~~(\mbox{SF,
one trap, fixed \emph{k}}).
\end{equation}
Interestingly, simulations show a distinct behavior for $m<3$, and
$m\geq 3$ (Figure \ref{fig2}). While in the case of $m\geq 3$ the
simulations agree with the theory (Eq. (\ref{sf_fixed_k})), as is
evident by the collapse of all curves with the same $kt$; for $m<3$
the decay of $\rho(t)$ is slower than exponential. Note that in
contrast to ER networks, $\rho(t)$ is larger for the denser networks
(smaller $\gamma$). Thus, whereas ER networks become less robust as
links are added, SF networks gain robustness. This is a fundamental
difference between ER and SF networks revealed by our results.

\begin{figure}
\includegraphics[width=7.5cm]{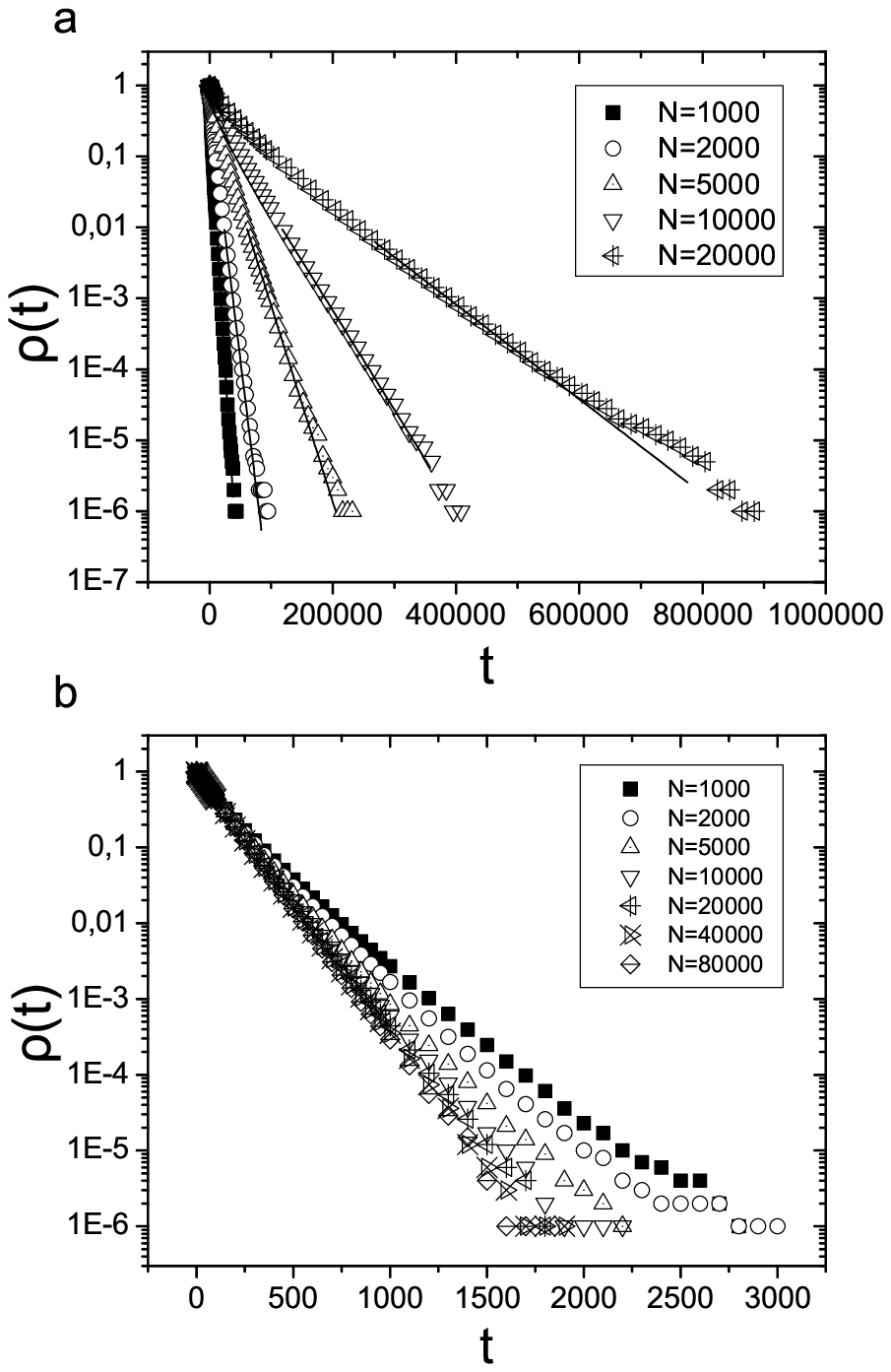}
\caption {Trapping in scale-free networks with $m=3$. (a) Particle
density $\rho(t)$ vs. $t$, for SF networks with $\gamma=2.5$, a
single trap on a random node, and different system sizes. Solid
lines represent fitting to exponential decay in the long time
regime, Eq. (\ref{sf_m3}). (b) Particle density $\rho(t)$ vs. $t$,
for SF networks with $\gamma=2.5$, traps with concentration $c=0.01$
placed on random nodes, and different system sizes.} \label{fig3}
\end{figure}

When the degree of the trap is allowed to vary, we consider the long
time regime. As in ER networks, the main contribution comes from
configurations in which the degree of trap is minimal i.e., $k=m$.
Thus we expect:
\begin{eqnarray}
\label{sf_m3} && \rho(t)\approx  \exp\left[-Amt/(N\av{k})\right]
\nonumber \\ && \mbox{(SF, one trap, random \emph{k}}, m\geq 3, t\gg
N\av{k})
\end{eqnarray}

which agrees with simulations (see Figure \ref{fig3}(a)). For SF
networks with many traps, a simple generalization of Eq.
(\ref{sf_m3}) (replacing $m$ by $nm$) is not applicable, and we
report only the numerical results (Figure \ref{fig3}(b)). Here,
similarly to ER networks, the smaller networks are more robust.

\begin{figure}
\includegraphics[width=7.5cm]{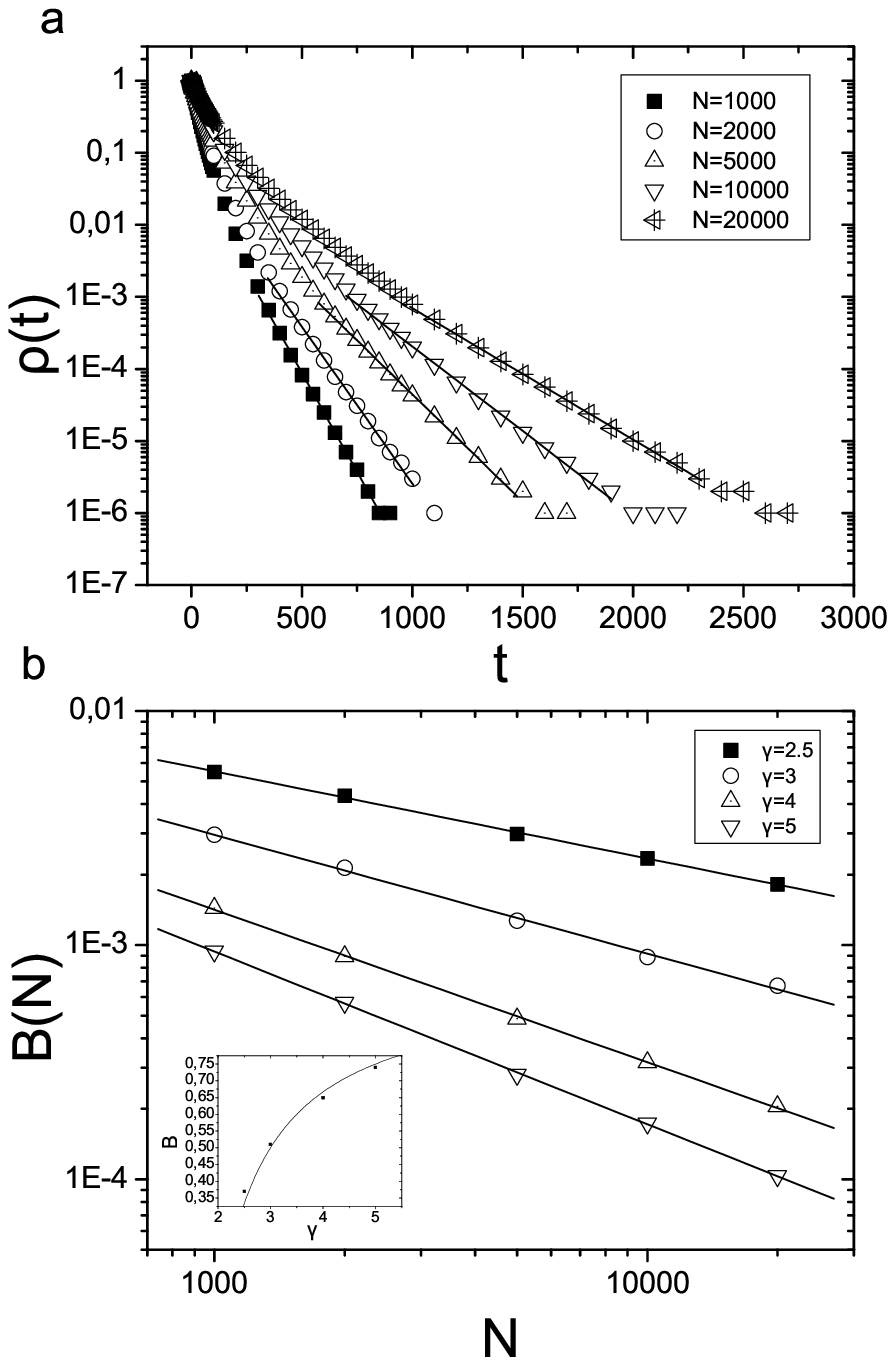}
\caption {Trapping in SF networks after failure of the most
connected node. (a) Particle density $\rho(t)$ vs. $t$, in SF
networks with $\gamma=2.5$ and $m=3$, for different system sizes.
One trap is placed on the node with the maximum degree. Solid lines
represent fitting to an exponential decay $\rho \sim e^{-Bt}$ in the
long-time regime. (b) The exponent $B$ vs. $N$, for $\gamma=2.5$
plotted in (a), as well as for $\gamma=3,4,5$. It can be seen that
$B\sim N^{-\beta}$ with $\beta \approx \frac{\gamma-2}{\gamma-1}$
(inset), in agreement with Eq. (\ref{sf_hub}).} \label{fig4}
\end{figure}

SF networks exhibit nodes of particular importance which have many
connections and play special role in transport \cite{Lopez}. Thus,
it is interesting to study a failure in the node of highest degree
(the hub) \cite{intentional}, which results in trapping of incoming
particles. The maximum degree $K$ in SF networks scales like $K
\approx mN^{\frac{1}{\gamma-1}}$ (for $\gamma>2$) \cite{resilience}.
Substituting $k=K$ in Eq. (\ref{sf_fixed_k}), we find:
\begin{eqnarray}
\label{sf_hub} &&\rho = \exp[-AtK/N\av{k}]\approx
\exp\left[-Amt/N^\frac{\gamma-2}{\gamma-1}\av{k}\right] \nonumber \\
&&\mbox{(SF, trap on the hub}, m\geq 3, t\gg N\av{k})
\end{eqnarray}
and the average time before trapping is thus $t_{\rm{tr}} \sim
N^{\beta}$ where $\beta=\frac{\gamma-2}{\gamma-1}<1$ (see Figure
\ref{fig4}). Realistic SF networks have $2<\gamma<3$
\cite{Barabasi_scaling,Albert_diameter,Faloutsos}, so that
$0<\beta<1/2$. This implies that real-world networks are ultra-prone
to failure in their highly connected nodes. This is an even stronger
effect compared to the targeted removal of high degree nodes
\cite{intentional}, whereas a failure of only one hub induces a
significant decrease in the trapping time, a finite concentration of
hubs has to be removed to fragment the network.

The dramatic decrease in the time before failure is not limited to
placing the trap precisely on the node with the maximal degree. It
can be proven that whenever we either \emph{(i)} choose
the node with maximal degree out of $n$ random nodes, when $n/N$ is
finite, or, \emph{(ii)} choose one of the $n$th nodes of highest
degree when $n=O(1)$, the probability of the trap degree to exceed
$K=mN^{\frac{1}{\gamma-1}}$ is finite. Thus, in these cases, the
trap will be attached to a sufficient number of links for the
scaling $t_{\rm{tr}} \sim N^{\frac{\gamma-2}{\gamma-1}}$ to appear.

The value of $\gamma$ for which SF networks are equivalent to ER
networks is a topic of recent interest \cite{cd}. Our results
suggest that SF networks are equivalent to ER only when $\gamma$ is
infinite, since only when $\gamma \rightarrow \infty$ does $\beta
\rightarrow 1$, as for homogenous ER networks. For ER networks, the
degree distribution is a Poisson with variance equals to the mean
$\av{k}$. Consequently, the typical maximal degree is roughly
$K\approx\av{k}+\sqrt{\av{k}}$. This yields $\rho\approx
\exp\left[-\frac{At}{N}\left(1+\frac{1}{\sqrt{\av{k}}}\right)\right]$,
such that the typical time is $t_{\rm{tr}}\sim N$ as before.

In the following, we study the behavior of the prefactor $A$. For
fully connected network and large $N$, $A \rightarrow 1$ (see, e.g.,
\cite{dani}). For sparse networks where the particle might be far
from the trap, $A$ is less than one, reflecting the fact that the
probability to follow a link to the trap is somewhat less than
$k_n/(N\av{k})$. To find the value of $A$, we first point out that
the trapping problem is a special case of a first passage time
problem \cite{Redner_book,disorderd_media,Noh} (since
$\rho(t)=1-\sum_{t'}^tF(t')$ where $F(t)$ is the probability to
reach the trap for the first time at time $t$). To calculate the
first passage time in networks, Baronchelli and Loreto \cite{ring}
used an approximate method that exploits the small-world nature of
most networks \cite{Albert_diameter,WS}. In theory, using the
adjacency matrix one can calculate the transition probability matrix
of the random walker, from which the first passage time can be
easily obtained via consecutive powers of the matrix (see below).
However, this is not feasible for large networks, and thus the
original random walk process was reduced to a random walk between
the network layers \cite{ring}. Given the trap, the number of nodes
$n_{\ell}$ that are in distance $\ell$ from it is calculated. Then,
a matrix $B$ of size $\ell_{\rm{max}} \times \ell_{\rm{max}}$ is
constructed, in which $B_{\ell,\ell'}$ is the probability of a
random walker in layer $\ell$ to jump into a node in layer $\ell'$.
For most real and model networks, $\ell_{\rm{max}} \sim \log{N}$
such that the size of the problem is reduced exponentially. Define
the number of links that connect layers $\ell$ and $\ell+1$ by
$s_{\ell}$, the number of links within layer $\ell$ by $o_{\ell}$,
and the sum of degrees of nodes in layer $\ell$ by
$m_{\ell}=(s_{\ell}+s_{\ell-1}+2o_{\ell})$. Since the random walker
jumps into each link with equal probability, the only non-zero
elements are: $B_{\ell,\ell+1}=s_{\ell}/m_{\ell}$,
$B_{\ell,\ell-1}=s_{\ell-1}/m_{\ell}$, and
$B_{\ell,\ell}=2o_{\ell}/m_{\ell}$. To represent the trap,
$B_{0,\ell}=0$ for all $\ell$ (since the trap forms layer $\ell=0$).
Since the probability for a random particle to start in a node of
layer $\ell$ is $n_{\ell}/(N-1)$, the first passage time probability
$F(t)$ is given by \cite{ring}:
\begin{equation}
\label{F}
F(t)=\sum_{\ell=1}^{\ell_{\rm{max}}}\frac{n_{\ell}}{N-1}(B^t)_{\ell,0}.
\end{equation}
In \cite{ring}, the matrix $B$ was constructed for ER networks, and
it was found that $F(t)\propto \exp\left[-Akt/N\av{k}\right]$, with
the value of $A$ determined numerically. In the following, we extend
this approach to study random regular and SF networks which were not
previously studied, and the behavior of $A$ for a concentration of
traps and different average degrees in ER networks.

To construct the transition probability matrix for $n$ traps in ER
networks, one has to calculate the number of nodes within distance
$\ell$ from \emph{each} of the $n$ nodes. This is easily
accomplished by setting $n_0=n$ and using the same formulae as in
\cite{ring}
$n_{\ell+1}=\left(N-\sum_{k=0}^{\ell}n_k\right)[1-(1-p)^{n_{\ell}}]$,
with $p=\av{k}/(N-1)$ being the independent probability of a link to
exist. The number of links connecting the different layers is the
same as in \cite{ring}:
$s_{\ell}=n_{\ell}\left(N-\sum_{k=0}^{\ell}n_k\right)p$, and
$o_{\ell}=n_{\ell}(n_{\ell}-1)p/2$, from which the matrix $B$ is
determined. We then calculate $F(t)$ from Eq. (\ref{F}), and $A$ by
fitting $F(t)$ to an exponential. We find that the change in $A$ for
$n>1$ is minor (of the order of ${\cal O}(N^{-1})$) and proportional
to $n$: $A(n)-A(n=1)\propto n$.

\begin{figure}
\includegraphics[width=8cm]{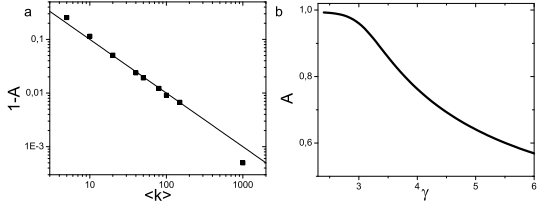}
\caption{An analysis of the prefactor $A$. (a) Plot of $1-A$ vs.
$\av{k}$ in ER networks, where $A$ is the fitting parameter in
Eq.~(\ref{rho_ER_1}). Here $N=10000$ and one trap was placed
randomly (black squares). The solid line corresponds to the
theoretical result $A = 1-1/\av{k}$ derived in the text. (b) $A$ vs.
$\gamma$ for SF networks with $N=10^8$ and $m=3$, when the trap is
located at the hub (theory only).} \label{fig5}
\end{figure}

We next study the dependence of $A$ on the average degree (for a
single trap). Applying the above method for ER networks with
different average degrees, we find that $A \approx 1-1/\av{k}$. This
is also confirmed by simulations (Figure \ref{fig5}(a)). For random
regular networks, we derived the transition matrix $B$,
from which we found that $A \approx 1-1/(k-1)$. Curiously, both
results can be written as $A \approx 1-1/(\kappa-1)$, where
$\kappa-1\equiv \av{k^2}/\av{k}-1$ is the branching factor of the
network (since in ER networks $\kappa=\av{k}+1$, and in regular
networks $\kappa=k$). A qualitative explanation of this relation
(which can also be recast as $A\approx 1-p_c$, where $p_c$ is the
percolation threshold \cite{resilience}) is still lacking.

For SF networks, we calculate $A$ for the case when the trap is
located at the node of maximal degree, by using \cite{tomography}
for the number of nodes in layer $\ell$ ($n_{\ell}$), and the number
of links emanating from layer $\ell$ into itself ($o_{\ell}$) and
into layer $\ell+1$ ($s_{\ell}$). As before, construction of the
transition matrix $B$, application of Eq. (\ref{F}), and fitting to
an exponential are used to calculate $A$ for different values of
$\gamma$ (Figure \ref{fig5}(b)). Using this method we can predict
$A$ for very large $N$s in which simulations are not possible. Here
the relation $A \approx 1-1/(\kappa-1)$ \cite{resilience} is valid
up to $\gamma \lesssim 4$.

\section{Conclusions}
We study the trapping problem on regular, ER, and SF networks using
theory and simulations. We develop a simple theory to account for
the behavior of the survival probability in a variety of conditions.
In ER networks we find that the trapping process exhibits a
non-exponential behavior which depends on both the number of traps
and the size of the network. For SF networks we find anomalous
behavior for networks with small minimal degree, expressed as
deviations from the theory. We also find that as opposed to ER
networks, particles survive for longer times in denser SF networks.
Finally, when the trap is placed in one the network hubs, we find a
new scaling with the system size. The average time before trapping
decreases dramatically in comparison to random failure or to ER
networks. This is true for all values of $\gamma$, suggesting that
the equivalence of SF and ER networks for $\gamma>4$ does not exist
for the trapping problem.

\section{Appendix 1: Finite probability for a choice of high degree traps}
\label{kmax}

We prove the following. Given a scale-free network with
$N\rightarrow \infty$ nodes and degree distribution
$P(k)=m^{\gamma-1}(\gamma-1)k^{-\gamma}~~,~~k\geq m$:
\subsection{Theorem}
\begin{enumerate}
\item The probability that the node with maximal degree out of $n\equiv cN$
random nodes will have degree that exceeds $mN^{\frac{1}{\gamma-1}}$
is finite, provided that $c=n/N$ is finite.
\item The probability that the degree of the node of $n$th largest degree
will exceed $mN^{\frac{1}{\gamma-1}}$ is finite, provided that
$n=O(1)$.
\end{enumerate}

\subsection{Proof}
\begin{enumerate}
\item
The probability of the node with maximal degree out of $n$ random
nodes to have degree $K$ is ${\cal P}(K)\approx
P(K)\left[\int_m^{K}P(k')dk'\right]^{cN}$. Substituting
$P(k)=cN(\gamma-1)m^{\gamma-1}k^{-\gamma}$ we have:
\begin{equation}
{\cal
P}(K)=cN(\gamma-1)m^{\gamma-1}K^{-\gamma}\left[1-(K/m)^{1-\gamma}\right]^{cN}.
\end{equation}
The probability for $K$ to be at least $mN^{\frac{1}{\gamma-1}}$ is:
\begin{eqnarray}
P\{K>mN^{\frac{1}{\gamma-1}}\} &=& \int_{mN^{\frac{1}{\gamma-1}}}^{\infty}{\cal P}(K)dK \nonumber \\
&=& \int_{mN^{\frac{1}{\gamma-1}}}^{\infty}cN(\gamma-1)m^{\gamma-1}K^{-\gamma}\left[1-(K/m)^{1-\gamma}\right]^{cN}dK \nonumber \\
&\geq&
cN(\gamma-1)m^{\gamma-1}\int_{mN^{\frac{1}{\gamma-1}}}^{\infty}K^{-\gamma}\left[1-(mN^{\frac{1}{\gamma-1}}/m)^{1-\gamma}\right]^{cN}dK
\nonumber \\ &=&
cN(\gamma-1)m^{\gamma-1}\int_{mN^{\frac{1}{\gamma-1}}}^{\infty}K^{-\gamma}e^{-c}dK
\nonumber \\ &=&
cN(\gamma-1)m^{\gamma-1}e^{-c}\left[K^{1-\gamma}/(1-\gamma)\right]_{K=mN^{\frac{1}{\gamma-1}}}^{\infty}
\nonumber \\ &=& ce^{-c}.
\end{eqnarray}
Thus, $P\{K>mN^{\frac{1}{\gamma-1}}\} \geq ce^{-c}$.
Since $c$ is finite, in every choice of $cN$ nodes there is a finite
probability that at least one of them will have degree larger than
$mN^{\frac{1}{\gamma-1}}$.

\item The probability of the node with $n$th largest degree in the
network to have degree $K$ is ${\cal P}(K)={N\choose n}n
P(K)\left[\int_K^{\infty}P(k')dk'\right]^{n-1}\left[\int_m^{K}P(k')dk'\right]^{N-n}$.
Inserting the degree distribution we have:
\begin{equation}
{\cal P}(K)={N\choose
n}n(\gamma-1)m^{\gamma-1}K^{-\gamma}\left[(K/m)^{1-\gamma}\right]^{n-1}\left[1-(K/m)^{1-\gamma}\right]^{N-n}.
\end{equation}
The probability for $K$ to be at least $mN^{\frac{1}{\gamma-1}}$ is:
\begin{eqnarray}
P\{K>mN^{\frac{1}{\gamma-1}}\} &=& \int_{mN^{\frac{1}{\gamma-1}}}^{\infty}{\cal P}(K)dK \nonumber \\
&=& {N\choose
n}n\int_{mN^{\frac{1}{\gamma-1}}}^{\infty}(\gamma-1)m^{\gamma-1}K^{-\gamma}\left[(K/m)^{1-\gamma}\right]^{n-1}\left[1-(K/m)^{1-\gamma}\right]^{N-n}dK \nonumber \\
&\geq& {N\choose
n}nm^{n(\gamma-1)}(\gamma-1)\int_{mN^{\frac{1}{\gamma-1}}}^{\infty}K^{n(1-\gamma)-1}\left[1-(mN^{\frac{1}{\gamma-1}}/m)^{1-\gamma}\right]^{N}dK \nonumber \\
&=& {N\choose
n}nm^{n(\gamma-1)}(\gamma-1)\int_{mN^{\frac{1}{\gamma-1}}}^{\infty}K^{n(1-\gamma)-1}e^{-1}dK \nonumber \\
&=& {N\choose
n}nm^{n(\gamma-1)}(\gamma-1)e^{-1}\left[K^{n(1-\gamma})/\left[n(1-\gamma)\right]\right]_{K=mN^{\frac{1}{\gamma-1}}}^{\infty}
\nonumber \\ &=& {N\choose n}e^{-1}N^{-n} \rightarrow e^{-1}(e/n)^n,
\end{eqnarray}
\end{enumerate}
where in the last step we used Stirling's approximation. Thus, if
$n=O(1)$, the probability that the node with $n$th largest degree is
greater than $mN^{\frac{1}{\gamma-1}}$ is finite.

\section{Appendix 2: Distances distribution in random regular networks}
\label{regular_section}

We derive the following results for the distribution of distances
from a random node in random regular networks. We imagine a process
in which all nodes have initially $k$ open links, and as the
algorithm proceeds, we connect open links to form the network edges.
At each step, we connect the open links emanating from layer $\ell$
into randomly open links from nodes which are outside layers
$1,...\ell$ to form layer $\ell+1$ (see \cite{tomography}). Denote
the number of nodes in distance $\ell$ from the root as $n_{\ell}$.
Define $r_{\ell}\equiv
1-\left[\sum_{\ell'=0}^{\ell}n_{\ell'}\right]/N$ as the fraction of
nodes outside layers $1,...,\ell$. Define also $s_{\ell}$ as the
number of edges connecting nodes in layer $\ell$ to nodes in layer
$\ell+1$, $o_{\ell}$ as the number of edges connecting nodes within
layer $\ell$, and $e_{\ell}$ as the number of edges emanating from
layer $\ell$ except for the edges incoming from layer $\ell-1$. The
following recursion relations hold:
\begin{eqnarray}
n_{\ell+1}&=&Nr_{\ell}\left[
1-\prod_{j=0}^{e_{\ell}-1}\left(1-\frac{k}{e_{\ell}-j+Nkr_{\ell}}\right)
\right] \\ r_{\ell+1} &=& r_{\ell}-n_{\ell+1}/N \\
e_{\ell+1} &=& kn_{\ell+1}-s_{\ell}
\\
o_{\ell+1} &=& \frac{e_{\ell+1}^2}{2(kNr_{\ell+1}+e_{\ell+1})} \\
s_{\ell+1} &=& e_{\ell+1}-2o_{\ell+1},
\end{eqnarray}
with $n_1=1$, $r_1=1-1/N$, $e_1=k$, $s_1=k$.

%
%
%
%
%
%
%

\bibliography{final_trapping}
\bibliographystyle{eplbib}

\end{document}